\documentclass[aps,prb,twocolumn,groupedaddress,showpacs,amsmath]{revtex4}
\usepackage{graphicx,amssymb}

\begin{document}

%

\title{Quantum tunneling of the N\'eel vector in antiferromagnetic [3~$\times$~3] grid molecules}

\author{O. Waldmann}
\email[Electronic address: ]{waldmann@iac.unibe.ch} \affiliation{Department of Chemistry and Biochemistry,
University of Bern, CH-3012 Bern, Switzerland}

\date{\today}

\begin{abstract}
Based on numerical calculations it is shown that the antiferromagnetic grid molecule Mn-[3~$\times$~3] is a
very promising candidate to experimentally detect the phenomenon of quantum tunneling of the N\'eel vector.
\end{abstract}


\maketitle

%

Molecular nanomagnets such as Mn$_{12}$ and Fe$_8$ have attracted much interest as they show quantum
tunneling of the magnetization at low temperatures.\cite{Mn12_Fe8} For antiferromagnetic (AF) molecular
wheels, a class of magnetic molecules in which an even number of ions is arranged in a ringlike
fashion,\cite{FWs} a different scenario associated with quantum tunneling of the N\'eel vector (QTNV) was
predicted theoretically,\cite{AC_NVT} but so far could not be observed experimentally. The identification of
a magnetic molecule which enables the detection of QTNV would represent a major achievement, of interest
also for applications in quantum technologies.

The experimental task to detect QTNV in molecular wheels is challenging: high sensitivity, low temperatures
and strong magnetic fields are mandatory. A technique such as electron spin resonance (ESR) could meet these
requirements, but does not couple to the N\'eel vector because of symmetry reasons.\cite{FM_NVT,OW_MQT_FW}
One way out would be to have a "sensor spin" which is coupled to only one of the AF
sublattices.\cite{FM_NVT} Its dynamics then will reflect the dynamics of the N\'eel vector. In principle, the
nuclear spin of a metal center or a nearby proton might act as a sensor spin. Another solution would be to
replace one of the magnetic ions on the ring by a dopant ion with different spin.\cite{FM_MFW} The resulting
excess spin also reflects the dynamics of the N\'eel vector - with the advantage of being detectable by ESR.
Efforts in this direction are under way, but the synthesis of appropriate modified wheels is quite difficult.

\begin{figure}
\includegraphics{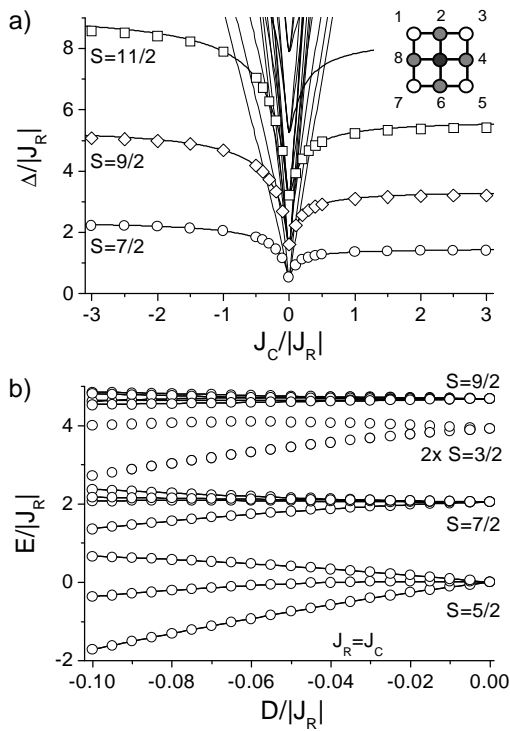}
\caption{\label{fig1}(a) Energies of the low-lying states with respect to the $S = 5/2$ ground state as
function of $J_C$ for $H^{3\times3}_{eff}$ (lines) and $H$ (symbols) with $D_R=D_C=0$ in zero field. For $H$,
only the energy gaps for the lowest states with $S = 7/2, 9/2, 11/2$ are displayed. The inset shows the
structure of the Mn-[3~$\times$~3] grid schematically. (b) Energy spectrum for $H^{3\times3}_{eff}$ (lines)
and $H$ (circles) as function of $D \equiv D_R = D_C$ for $J_C=J_R$ at zero field. The spin multiplets get
split by the anisotropy. The $S=5/2$ level for instance split in the sequence $|M| = 1/2, 3/2, 5/2$. Twofold
degenerate spin levels such as the $S = 3/2$ levels at about $3.5 |J_R|$ are not reproduced by
$H^{3\times3}_{eff}$. They belong to an $E$ band similar as in wheels,\cite{OW_FW_DYN,OW_Cr8} but are not
relevant here as explained in the text.}
\end{figure}

Recently, a supramolecular Mn-[3~$\times$~3] grid was investigated.\cite{OW_Mn3x3,OW_Mn3x3_II} Here, nine
spin-5/2 Mn(II) ions occupy the positions of a regular 3\:$\times$\:3 matrix, held in place by a lattice of
organic ligands [see inset of Fig.~\ref{fig1}(a)]. The unusual magnetic properties suggested that the
[3~$\times$~3] grid can be regarded as an AF "ring" doped with a central Mn ion.\cite{OW_Mn3x3} Indeed, the
spin Hamiltonian of a [3~$\times$~3] grid, which for an idealized structure reads \cite{Dipdip}
\begin{eqnarray}
\label{H3x3}
 H = && -J_R \left( \sum^7_{i=1}{ \textbf{S}_i \cdot \textbf{S}_{i+1} }
 + \textbf{S}_8 \cdot \textbf{S}_1 \right) + D_R \sum^8_{i=1} S^2_{i,z}
 \cr
 && -J_C \left( \textbf{S}_2 + \textbf{S}_4 + \textbf{S}_6 + \textbf{S}_8 \right) \cdot \textbf{S}_9
 \cr
 &&  + D_C S^2_{9,z} + g \mu_B \textbf{S} \cdot \textbf{B},
\end{eqnarray}
consists of a part involving only the ring of the eight peripheral metal ions, a part related to the central
ion, and terms representing an interaction between these two sets of spin centers [$\textbf{S}_i$ is the
spin operator of the $i$th ion with spin $s = 5/2$, spins at "corners" are numbered 1, 3, 5, 7, those at
"edges" 2, 4, 6, 8, and the central spin is numbered 9, see Fig.~\ref{fig1}(a); $z$ denotes the axis
perpendicular to the grid plane]. For Mn-[3~$\times$~3], an AF intraring coupling ($J_R<0$) and uniaxial
easy-axis anisotropy ($D_R<0$) was found.\cite{OW_Mn3x3,OW_Mn3x3_II} Furthermore, the central spin is
coupled to only one of the sublattices of the peripheral ring, namely to the edge spins. Thus, magnetically
the Mn-[3~$\times$~3] grid should correspond to an AF ring of eight spin-5/2 centers with a sensor spin
naturally built in. This work demonstrates that Mn-[3~$\times$~3] is indeed a promising molecule to detect
QTNV experimentally.

%

The dimension of the Hilbert space is huge for Mn-[3~$\times$~3]. However, the low-energy sector of an AF
ring (with a small, even number of sites $N$) is very well described by approximating the wave functions by
$|\beta_A \beta_B S_A S_B S M \rangle$ with $S_A = S_B = Ns/2$.\cite{OW_MQT_FW,OW_FW_DYN,AH_NVT} Here, $S_A$
($S_B$) denotes the total spin of sublattice $A$ ($B$), and $\beta_A$, $\beta_B$ abbreviate intermediate
quantum numbers (omitted in the following). Physically, this approach works well because the internal spin
structure due to the dominant Heisenberg interaction is essentially classical.\cite{OW_Cr8} Applying this
approximation to the ring terms in Eq.~(\ref{H3x3}), an effective Hamiltonian for the Mn-[3~$\times$~3] grid
is obtained:
\begin{eqnarray}
\label{Heff}
 H^{3\times3}_{eff} = && - \tilde{J}_R \textbf{S}_A \cdot \textbf{S}_B
 + \tilde{D}_R ( S^2_{A,z} + S^2_{B,z} )
 \cr
 && - J_C \textbf{S}_A \cdot \textbf{S}_9  + D_C S^2_{9,z} + g \mu_B \textbf{S} \cdot \textbf{B}.
\end{eqnarray}

Here, $\tilde{J}_R = 0.526 J$ and $\tilde{D}_R = 0.197 D_R$. $A$ ($B$) denotes the sublattice of corner
(edge) spins. Figure~\ref{fig1} compares results calculated for $H^{3\times3}_{eff}$ and $H$. The agreement
is quite good. $H^{3\times3}_{eff}$ is designed to reproduce the lowest-lying states optimally, higher-lying
states are thus increasingly less well described.

Remarkably, $H^{3\times3}_{eff}$ works well in a wide range of $J_C/|J_R|$ values. In the strong coupling
limit, $|J_C| \gg |J_R|$, the wave functions should be approximated by $|S_B S_9 S_{B9} S_A S M\rangle$,
i.e., by coupling first the spins $S_B$ and $S_9$, and then the resulting spin $S_{B9}$ with $S_A$. In the
weak coupling limit, however, the appropriate wave functions would be $|S_A S_B S_{AB} S_9 S M\rangle$, where
first $S_A$ and $S_B$ are coupled together. As both wave functions lead to the same effective Hamiltonian,
Eq.~(\ref{Heff}), a broad range of $J_C/|J_R|$ is covered. This suggests a general strategy to construct an
effective Hamiltonian: Sublattices as expected from the classical spin configuration are introduced, and then
spins of each sublattice are replaced by the mean-field spin ${\bf S}_{SL} = 1/N_{SL} \sum_{i\in SL} {\bf
S}_i$ ($N_{SL}$ denotes the number of spins of sublattice $SL$). The effect of weak quantum fluctuations may
be accounted for by renormalizing the parameters of the effective Hamiltonian, hence $\tilde{J}_R$ and
$\tilde{D}_R$ in Eq.~(\ref{Heff}).\cite{OW_MQT_FW}

Interestingly, a similarity of the single-molecule magnet Mn$_{12}$ \cite{Mn12_Fe8,NR_Mn12} and a
[3~$\times$~3] grid turns up: Mn$_{12}$ may be regarded as an octanuclear ring of Mn(III) ions doped by a
central Mn(IV) tetramer. The above considerations then lead directly to an effective Hamiltonian, which, in
view of the success of this approach for the molecular wheels and the [3~$\times$~3] grid, is expected to
describe also the relevant states of Mn$_{12}$ well.

%

With $H^{3\times3}_{eff}$, QTNV in Mn-[3~$\times$~3] can now be readily analyzed numerically. The energy
spectrum as function of magnetic field is shown in Fig.~\ref{fig2} for several field orientations and
$J_R=J_C=-5$~K, $D_R=D_C=-0.14$~K ($\theta$ denotes the angle between magnetic field and $z$ axis; the
azimuthal angle $\varphi$ has no effect, as is evident from $H$, $H^{3\times3}_{eff}$). These parameters are
supported by recent inelastic neutron scattering experiments,\cite{Mn3x3_ins} and torque measurements at very
low temperatures.\cite{OW_Mn3x3_II} The semiclassical tunneling action $S_0/\hbar = N S_i \sqrt{2 D_R/J}$
(QTNV sets in for $S_0/\hbar \gg 1$, see Ref.~\onlinecite{AC_NVT}) is then estimated to $S_0/\hbar = 4.7$,
which is even larger than for Fe$_{10}$ ($S_0/\hbar = 3.3$) and CsFe$_8$ ($S_0/\hbar =
3.9$).\cite{OW_MQT_FW,FM_MFW} This demonstrates that Mn-[3~$\times$~3] is indeed a favorable candidate with
respect to QTNV. Some following points should be noted:

\begin{figure}
\includegraphics{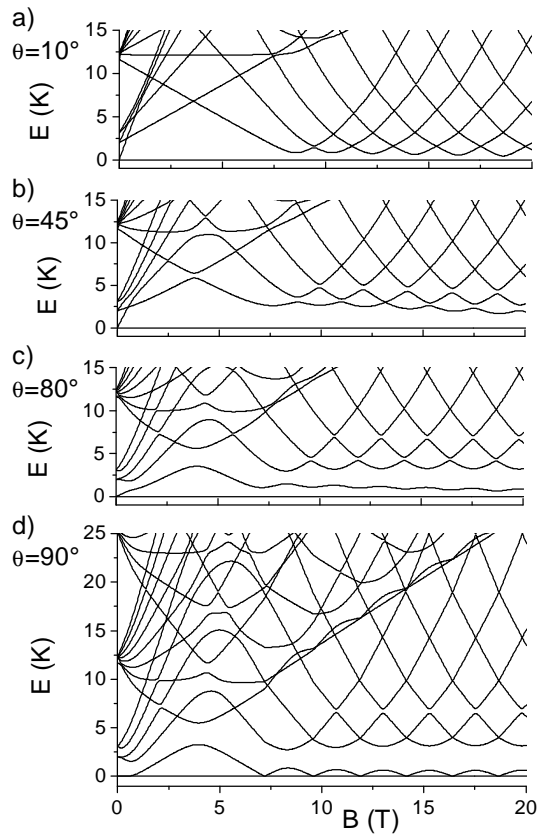}
\caption{\label{fig2} Energy spectrum vs magnetic field for different orientations $\theta$ of the magnetic
field (the energy of the lowest state was set to zero at each field). For zero field, the states with energy
below 3~K belong to the $S = 5/2$ ground state multiplet (which exhibits a hard-axis zero-field splitting),
those at $\approx$ 12.5~K to the next higher $S$ = 7/2 multiplet, and those at $\approx$ 25~K to the $S$ =
9/2 multiplet.}
\end{figure}

1) Concerning the dependence on $J_C/|J_R|$, the states should be divided into the energetically lowest
states for $S=5/2, 7/2, 9/2, \ldots$, and the remainder of the states (in the terminology of
Refs.~\onlinecite{OW_FW_DYN} and \onlinecite{OW_Cr8}, these states form the $L$ band, while the remaining
states belong to either the $E$ band or the quasicontinuum). Obviously, only the first set of states (the
$L$ band) is relevant for the low-lying part of the energy spectrum in magnetic fields. As soon as $J_C$
assumes significant values, the relative energy spacings of these states, and therewith the relative field
positions of the ground-state level crossings, are close to those found in the strong-coupling limit $|J_C|
\gg |J_R|$, even if $|J_C|$ is much smaller than $|J_R|$. The weak-coupling limit is realized only for
$|J_C/J_R| \lesssim 0.002$. Thus, the results shown in this work for $J_C = J_R$ are actually characteristic
for the strong coupling limit.

2) The energy gap between ground and first excited state, $\Delta(B)$, shows a $|\sin(\pi g \mu_B B / J_R +
\alpha)|$-like oscillatory field dependence for high fields in the grid plane [Fig.~\ref{fig2}(d)]. Such a
behavior was identified as a characteristics of QTNV in the (modified) molecular wheels.\cite{AC_NVT,FM_MFW}
$\alpha$ accounts for a shift in field.

3) Finally, $\Delta(B)$ periodically drops to zero indicating level crossings (LCs) at fields $B_m$ ($m \in
\mathbb{N}$).\cite{OW_MQT_FW,AH_NVT} Such LCs trivially exist for $\theta=0^\circ$. However, for canted
magnetic fields gaps appear at each LC field, i.e., the ground state LCs become anticrossings signaling
level mixing (Fig.~\ref{fig2}). Starting with $\theta=0^\circ$, the gaps at the LC fields first increase,
reach maximal values at around $\theta=45^\circ$, and then shrink to disappear again at $\theta=90^\circ$. In
the (unmodified) wheels, $\Delta(B_m)=0$ is enforced by symmetry for all field
directions.\cite{OW_MQT_FW,AH_NVT} This symmetry is absent in [3~$\times$~3] grids, and for fields with
$\theta \neq 0^\circ$ gaps open at the LCs due to the action of the uniaxial magnetic anisotropy: In a
coordinate frame with $z'$ parallel to the field, one obtains $D S^2_z = D' S'^2_z + E' (S'^2_x-S'^2_y) - G'
(S'_x S'_z + S'_z S'_x)$. The $G'$ term mixes the two levels involved in a LC (for which $|\Delta M'|=1$),
leading to the above behavior as $G' = \sin(\theta) \cos(\theta) D /3$. This angular behavior of $\Delta(B)$
has important implications for the QTNV scenario (\emph{vide infra}).

%

\begin{figure}
\includegraphics[scale=0.95]{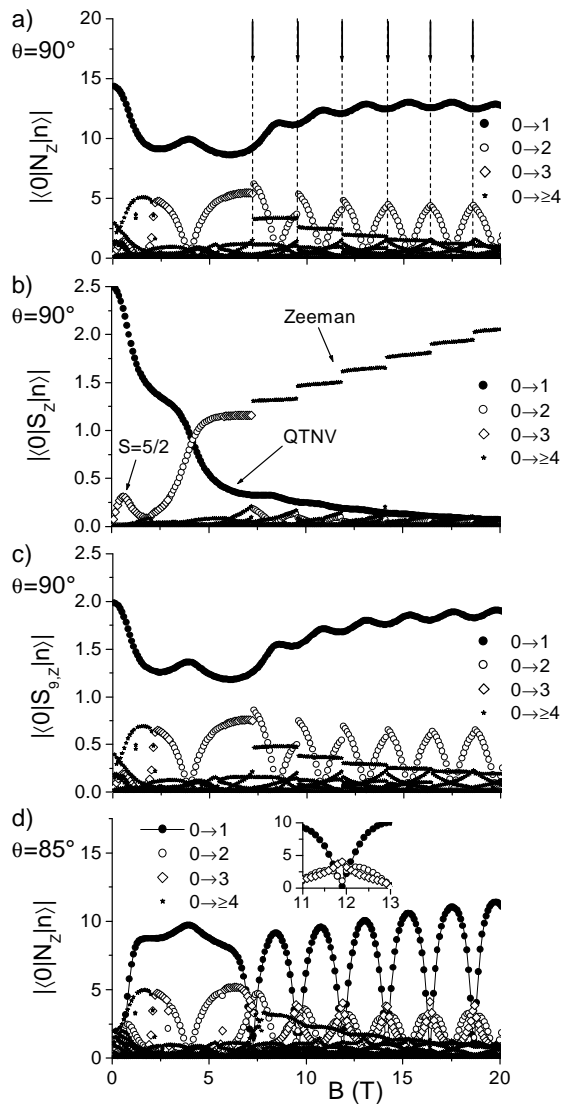}
\caption{\label{fig3} Absolute matrix elements between the ground and the 20 next higher-lying states for
the $z$ component of (a) the N\'eel vector, (b) the total spin, and (c) the central spin for fields in the
plane of the grid. Arrows and dashed lines in panel (a) indicate the positions of LCs. (d) Matrix elements
for the N\'eel vector with slightly canted field. The inset magnifies the field range 11 - 13\,T for the
first three matrix elements.}
\end{figure}

In order to analyze QTNV, the matrix elements between the ground state $|0\rangle$ and the 20 next higher
lying states $|n\rangle$ were calculated for the $z$ component of the N\'eel vector $|\langle 0 | N_z | n
\rangle|$, the total spin $|\langle 0 | S_z | n \rangle|$, and the central spin $|\langle 0 | S_{9,z} | n
\rangle|$ ($n=1, \ldots, 20$ numbers the states by their energy at each field). The N\'eel vector was
defined as ${\bf N}={\bf S}_A-{\bf S}_B+{\bf S}_9$. The results as function of magnetic field are shown in
Figs.~\ref{fig3}(a)-(c) for $\theta=90^\circ$.

For the N\'eel vector, the excitation between ground and first excited state, $|\langle 0 | N_z | 1
\rangle|$, which corresponds to QTNV, is dominant [Fig.~\ref{fig3}(a)]. The energy gap $\Delta(B)$ is thus
identified as the tunneling gap due to QTNV (for fields beyond the first LC). Higher-lying excitations, in
particular, to the second excited state, are also present. Accordingly, the correlation function vs time,
$\langle 0 | N_z(t) N_z(0) | 0 \rangle \propto \sum_n |\langle 0 | N_z | n \rangle|^2 \exp(-{i \over \hbar}
E_n t)$, shows a dominant oscillation at the tunneling frequency $\Delta/\hbar$ disturbed by higher frequent
oscillations. As the contribution of higher-lying excitations is minimal for magnetic fields inbetween LCs
[where the tunneling gap $\Delta(B)$ assumes maxima, see Fig.~\ref{fig2}(d)], the dynamics of the N\'eel
vector is best described by QTNV at these fields.

In $S_z$, three excitations have significant intensities [Fig.~\ref{fig3}(b)]. The excitation with maximum at
$\approx 1$\,T corresponds to a transition within the $S=5/2$ ground-state multiplet. The excitation gaining
increasing intensity from $\approx 5$\,T onwards corresponds to a transition to the Zeeman-split level, which
is visible in the energy spectrum at energies $\approx g \mu_B B$ [Fig.~\ref{fig2}(d)]. And finally, the
excitation with an approximate $B^{-2}$ field dependence corresponds to the transition between the ground and
first excited state reflecting exactly QTNV.\cite{FM_MFW}

The matrix elements for the central spin basically just behave as those of the N\'eel vector
[Fig.~\ref{fig3}(c)]. This is of course expected, but demonstrates unambiguously that the central spin of a
[3~$\times$~3] grid indeed acts as a sensor spin for the dynamics of the spins on the peripheral ring.

The results resemble those for molecular wheels: The field dependencies of $\Delta(B)$, $|\langle 0 | S_z | 1
\rangle|$, $|\langle 0 | N_z | 1 \rangle|$ (and $|\langle 0 | N_z | 2 \rangle|$) show similar behavior as for
the modified and unmodified wheels.\cite{AH_NVT,FM_MFW} As shown above, the strong coupling limit is
effectively realized in Mn-[3~$\times$~3]. Then the sublattice spin $S_B$ and the central spin $S_9$ act as a
combined larger spin $S_{B9}$, and $H^{3\times3}_{eff}$ simplifies to $H' = -J' \textbf{S}_A \cdot
\textbf{S}_{B9} + D'_A S^2_{A,z} + D'_B S^2_{B9,z} + g \mu_B \textbf{S} \cdot \textbf{B}$, which is exactly
the effective Hamiltonian of a modified wheel.\cite{FM_MFW} Thus, magnetically Mn-[3~$\times$~3] behaves like
a modified AF wheel.

The new feature observed here is that, along with the opening of gaps at the LCs, the behavior of the matrix
elements changes drastically for $\theta \neq 0,90^\circ$, Fig.~\ref{fig3}(d) (preliminary calculations
confirmed similar behavior for the modified wheels). At the LC fields, $|\langle 0 | N_z | 1 \rangle|$ (and
$|\langle 0 | N_z | 2 \rangle|$) vanishes and $|\langle 0 | N_z | 3 \rangle|$ becomes the dominant matrix
element. Apparently, the opening of gaps at the LCs and the rise of $|\langle 0 | N_z | 3 \rangle|$ results
in a breakdown of the QTNV scenario at these fields. Thus, the picture of QTNV needs an extension near LCs in
the case of Mn-[3~$\times$~3] or modified wheels. However, importantly, in those regions of the magnetic
field where the tunneling gap $\Delta(B)$ is large, QTNV persists even for sizeable canting angles.

%

Finally, the prospects of an experimental observation of QTNV in Mn-[3~$\times$~3] is discussed. Having shown
that the central spin acts as sensor spin and the analogy with a modified wheel, the considerations of
Refs.~\onlinecite{FM_NVT},\onlinecite{FM_MFW} are applicable to Mn-[3~$\times$~3]. Only main points will be
addressed here, further details may be found in these works.

QTNV can be detected by the highly sensitive ESR technique, as is evident from Fig.~\ref{fig3}(b) noting that
the ESR intensity is proportional to $|\langle 0 | S_z | 1 \rangle|^2$ for $\theta=90^\circ$. For
Mn-[3~$\times$~3], the most favorable field regime would be 10-20~T, see Fig.~\ref{fig3}(a). In this regime,
the ESR intensity is smaller by two orders of magnitude compared to the maximal intensity at $B=0$, but is
still large enough to be detectable with today's ESR spectrometers, especially as rather large high-quality
single crystals are available.\cite{LK_private} The magnetic field should be oriented close to the plane of
the grid, but a canting of about 5$^\circ$ is well tolerable if QTNV is measured near the tops of the
tunneling gap $\Delta(B)$, i.e., at fields inbetween LCs. In a continuous-wave ESR experiment, the linewidth
of the signal provides an upper limit for the decoherence rate $\Gamma_S$, which allows one to test the
coherence condition $\Gamma_S < \Delta / \hbar$, or if QTNV in Mn-[3~$\times$~3] is coherent,
respectively.\cite{FM_NVT}.

Nuclear magnetic resonance (NMR) experiments are an alternative. As shown in Ref.~\onlinecite{FM_NVT}, QTNV
leads to an absorption peak in the NMR signal at $\omega = \Delta / \hbar$, but with an intensity
significantly reduced as $(A/\Delta)^2$ ($A$ is the hyperfine coupling constant). For molecular wheels it was
suggested to measure near LCs, where $\Delta$ is small, in order to gain intensity. But for
Mn-[3~$\times$~3], this would require an unrealistically accurate orientation of the magnetic field
perpendicular to the grid plane. Fortunately, the sensitivity of Mn$^{55}$-NMR is 3$\times$10$^5$ times
larger than of Fe$^{57}$-NMR,\cite{NMR_sens} resulting in detectable signals also near the maxima of
$\Delta(B)$ in Mn-[3~$\times$~3]. And, as mentioned already, relatively large single crystals are
available.\cite{LK_private} It is added that the NMR signal due to QTNV can be easily identified by its field
dependence which differs markedly from that of the Lamor frequency.

%

In conclusion, the molecular grid Mn-[3~$\times$~3] is an exciting prospect for observing quantum tunneling
of the N\'eel vector experimentally: It has a sensor spin naturally built in which enables a detection of
tunneling by ESR, and it features the highest value of the tunneling action $S_0/\hbar$ known to date. The
issue of coherence is of great importance for any potential applications. In view of tunneling gaps of 1~K,
coherent QTNV might be realized in Mn-[3~$\times$~3].

%
\begin{acknowledgments}
I thank F. Meier for many enlightening discussions on the subject.
\end{acknowledgments}

%

%

\begin{references}
\bibitem{Dipdip}
Intramolecular dipole-dipole interactions are not negligible, but have similar effects as the single-ion
anisotropy and are well grasped by $S^2_{i,z}$ terms. The $D$ values should be understood as to include both
anisotropy contributions.
\end{references}
\end{document}